\centerline{\bf Thermonuclear Supernovae: Simulations of the Deflagration Stage 
and Their Implications}

\bigskip

\centerline{V. N. Gamezo$^{1\ast}$, A. M. Khokhlov$^1$, E. S. Oran$^1$,
A. Y. Chtchelkanova$^2$, R. O. Rosenberg$^3$}
\vskip 1truecm

\centerline {$^1$Laboratory for Computational Physics and Fluid Dynamics,}
\centerline {Naval Research Laboratory, Washington, D.C. 20375, USA}
\medskip
\centerline {$^2$Strategic Analysis Inc., Arlington, VA 22201, USA}
\medskip
\centerline {$^3$Center for Computational Science,}
\centerline {Naval Research Laboratory, Washington, D.C. 20375, USA} 
\bigskip
\bigskip

\centerline{ $^\ast$To whom correspondence should be addressed.}
\centerline{ E-mail: gamezo@lcp.nrl.navy.mil}
\vskip 2truecm

\centerline{ {\sl Science} Express (online), November 21, 2002. 
To appear in {\sl Science}, January 2003 }

\vfill


\centerline{ \bf Abstract}
\bigskip

\noindent{\bf
Large-scale three-dimensional numerical simulations of the deflagration
stage of a thermonuclear supernova explosion show the formation and
evolution of a highly convoluted turbulent flame in a gravitational field
of an expanding carbon-oxygen white dwarf. The flame dynamics is dominated
by the gravity-induced Rayleigh-Taylor instability that controls the
burning rate. The thermonuclear deflagration releases enough energy to
produce a healthy explosion.  The turbulent flame, however, leaves large
amounts of unburnt and partially burnt material near the star center,
whereas observations imply these materials only in outer layers. This
disagreement could be resolved if the deflagration triggers a detonation.
}
\vfill\eject

\noindent
According to observations and models, many stars that stea\-di\-ly
burn their nuclear fuel for millions or billions of years suddenly end
their lives with a powerful explosion that produces a bright object
called a supernova. A supernova explosion can be powered either by the
gravitational energy released during the core collapse of a massive star,
or by the nuclear energy released by explosive thermonuclear burning of
a star. Here, we focus on thermonuclear supernovae that belong to the
Type Ia (SN~Ia) in the observation-based classification ({\sl1-3}).

Thermonuclear supernovae are produced by explosions of white dwarfs (WD),
small dense stars composed of carbon and oxygen nuclei and detached
degenerate electrons ({\sl1,3-10}). The term degenerate means that
electrons occupy all possible quantum states below a certain energy. The
hydrostatic equilibrium in a WD is supported for the most part by
the degenerate-electron pressure that does not depend on temperature.
White dwarfs form at the end of the evolution of stars whose original
masses are less than 8 solar masses ($M_\odot$). A star can lose a large
fraction of its material by ejecting outer layers into space at the final
stages of evolution.  The mass of a remaining WD is always less than the
Chandrasekhar limit, $1.4M_\odot$, above which a hydrostatic equilibrium
of degenerate matter is impossible.  An isolated carbon-oxygen WD is
stable and almost inert because its temperature is not high enough to
induce any substantial nuclear reactions.  This isolated dead star can
exist almost indefinitely, slowly cooling down as it radiates its energy
into space.  Observations show, however, that more than 50\% of all stars
are not isolated. They belong to groups of two or more stars that orbit
a common center of mass.  In a close binary system, a WD can increase
its own mass by accreting material from its companion star. Such systems
are considered to be the most probable SN~Ia progenitors, even though the
exact nature of the companion star and the details of the mass accretion
are still unclear ({\sl1,3-5,7-9}).

When the mass of the WD approaches the Chandrasekhar limit, any
small mass increase results in a substantial contraction of the star,
and the material near its center is compressed.  This increases the
temperature and accelerates thermonuclear reactions near the center.
The released energy further increases the temperature, thus further
accelerating thermonuclear reactions.  This process is slowed down
by the neutrino, convective and conductive cooling.  Nevertheless, the
temperature in the WD center rises and reaches the point where the energy
release overwhelms the energy outflow.  In an ordinary  non-degenerate
star, the energy release would be stabilized by a thermal expansion
accompanied by the work against gravity. In a WD, however, the initial
temperature increase does not affect the degenerate-electron pressure,
and, therefore, does not lead to any substantial expansion that could
slow down thermonuclear reactions and prevent the runaway process.
Eventually, the temperature increases to the level where the thermal
and the degenerate-electron pressure components become comparable,
and the material begins to expand. At that time, however, the expansion
is already unable to quench the fast thermonuclear burning ignited in
the center of a WD.  The thermonuclear runaway mechanism in degenerate
matter was first described in ({\sl11}) and now is a key component of
all plausible SN~Ia scenarios ({\sl1,4-9}). Depending on the scenario,
the ignition may occur near the center, off center, or in outer layers
of the star.  Here, we consider the central ignition mode.

Ignition starts a SN~Ia explosion that lasts only a few seconds,
but releases $\simeq10^{51}$ ergs, about as much energy as the Sun
would radiate during 8 billion years.  The energy is produced by
a network of thermonuclear reactions that begins from $^{12}$C and
$^{16}$O nuclei and ends in $^{56}$Ni and other iron-group elements.
Large amounts of intermediate-mass elements, such as Ne, Mg, Si, S, and
Ca, are created as well.  The main energy-producing reactions occur in a
thin layer, called a thermonuclear flame, that propagates outwards. At
the beginning, the flame is laminar and its propagation velocity is
controlled by thermal conductivity. As the flame moves away from the
center, it becomes turbulent and accelerates.  Eventually, the burning
can undergo a transition from a relatively slow, subsonic regime, called
deflagration, into a supersonic regime, called detonation, where the
reaction front is preceded by a shock wave.  Most of the energy released
during the explosion transforms into kinetic and thermal energies of
the expanding material.  When the sum of kinetic ($E_k$) and thermal
($E_t$) energies exceeds the potential energy of self-gravitation $E_g$,
the star becomes unbound, such that expanding material is no longer
bound by gravity and will continue to expand indefinitely.

Thermonuclear reactions that occur during the explosion provide energy for
the expansion, but not for the luminosity of the expanding gas observed
as a SN~Ia. The energy source for the luminosity is the slow radioactive
decay sequence $^{56}$Ni$ \rightarrow ^{56}$Co$ \rightarrow ^{56}$Fe.
The luminosity reaches its maximum 15 to 20 days after the explosion,
and then decreases slowly until all the $^{56}$Co decays. The maximum
brightness of a SN~Ia is comparable to the brightness of an entire
galaxy, and can vary by an order of magnitude from one supernova to
another. Observations also show that the maximum luminosity of SNe~Ia in
visible wavelengths correlates with the rate at which the luminosity
decreases after the maximum ({\sl12-14}). This and other approximate
correlations summarized in ({\sl2,8}) make it possible to use SNe~Ia as
standard candles to measure distances and estimate cosmological parameters
critical for our understanding of the global evolution of the Universe
({\sl4,5,7-10,15-19}).

The importance of SNe~Ia as standard candles grows as observational
techniques improve and estimations of cosmological parameters become more
accurate.  We are approaching the limit, however, where the reliance on
empirical correlations becomes the main source of uncertainty. Even though
the correlation between the maximum luminosity and rate of decrease of
the luminosity for SNe~Ia has a theoretical explanation ({\sl20}) based on
a one-dimensional model, this is at best only a first approximation that
does not take into account a detailed explosion mechanism and its possible
variations. The only way to solve this problem is to study details of
supernova explosions using multidimensional numerical simulations.

One-dimensional (1D) numerical models have been extensively
used to test general ideas about possible explosion mechanisms
({\sl21-24,4,6}).  Delayed-detonation models ({\sl25-32}), that postulate
a deflagration-to-detonation transition (DDT) at some stage of the
thermonuclear explosion, are most successful in reproducing observed
characteristics of SNe~Ia. Many important details, however, including
the mechanism of DDT, are still unknown because SN~Ia explosions are
intrinsically three-dimensional (3D) phenomena.

Only a full-scale 3D numerical model may be expected to reproduce
all key features of the explosion that involves propagation of a
turbulent thermonuclear flame in the gravitational field of a white
dwarf. Building such a model is a complicated interdisciplinary problem on
the leading edge of astrophysics, nuclear physics, combustion physics,
and computational physics. Full-scale 3D numerical simulations of
thermonuclear supernova explosions have become a reality during the
last few years ({\sl33-35}), in great part owing to the progress in
computational technology.  Here, we describe a 3D numerical model,
present numerical results for the deflagration stage of the explosion,
and discuss implications of the results for astrophysics of SNe~Ia.

\bigskip
\noindent{\bf Physical and Numerical Model.}
The numerical model described in more detail in ({\sl36}) is based on
reactive Euler equations that represent mass, momentum, and energy
conservation laws for an inviscid fluid. The thermodynamic properties
of the fluid are defined by the equation of state of degenerate matter,
which is well known from basic theory and includes contributions from
ideal Fermi-Dirac electrons and positrons, equilibrium Planck radiation,
and ideal ions. Reactions involved in the thermonuclear burning of the
carbon-oxygen mixture are described  by a simplified four-equation kinetic
scheme ({\sl25,33}).  This kinetic scheme is coupled with a flame-capturing
technique ({\sl37,33}) that introduces an additional partial differential 
equation and ensures thermonuclear flame propagation at a prescribed 
speed $S$.  The resulting set of equations is integrated on
a Cartesian adaptive mesh using an explicit, second-order, Godunov-type
numerical scheme ({\sl37,38}).  The mesh is dynamically refined around
shock waves, flame fronts, and in regions of steep gradients of density,
pressure, composition, and tangential velocity. The computational cell size
$dx$ varies within predefined limits $dx_{min}$ and $dx_{max}$.

The large-scale simulations described here do not resolve the physical
thickness of a flame that differs from the white dwarf radius $R_{WD}$
by up to 12 orders of magnitude. Therefore, the flame speed must be
provided by an additional subgrid model that takes into account physical
processes at scales smaller than the computational cell size. We define
the flame speed $S$ as
$$
S = \max ( S_l, S_t ),                            \eqno(1)
$$
where the steady-state laminar flame speed $S_l$ in carbon-oxygen
degenerate matter is a known function of temperature, density, and
composition ({\sl39,40}). The speed $S_t$ of a quasi-steady-state
turbulent flame driven by the gravity-induced Rayleigh-Taylor (RT)
instability depends on the gravitational acceleration $g$ and the length
scale $L$ ({\sl37,41}):
$$
S_t \simeq 0.5 \sqrt{AgL}                         \eqno(2)
$$
\noindent
where $A=(\rho_0 - \rho_1)/(\rho_0 + \rho_1)$ is the Atwood number, and
$\rho_0$ and $\rho_1$ are  the densities ahead and behind the flame front,
respectively.  The driving scale $L$ is set equal to one or two computational
cell sizes $dx$. 

Equation~(2) is based on the two main properties of a turbulent flame
({\sl37,41}): self-similarity of the flame structure and self-regulation
of the flame speed.  Self-similarity means that the 3D distortions of
the flame surface at different scales are similar.  Self-regulation
means that changing the flame speed at small scales does not affect
the flame speed at larger scales. This occurs because a higher flame
speed at small scales causes small flame wrinkles to burn out, thus
decreasing the flame surface.  The resulting burning rate, defined as
a product of the flame speed at small scales and the flame surface,
does not change. This subgrid model makes it possible to reproduce the
correct flame propagation in numerical simulations while explicitly
resolving only the large-scale flame structure.  If the resolved flame
structure is self-similar and self-regulating, and behaves according to
the Eq.~(2), the subgrid model just extends this behavior to unresolved
small scales.  Shifting the boundary between resolved and unresolved
scales by changing the numerical resolution should not affect the
turbulent flame propagation.  The solution obtained should then be
independent of numerical resolution and on the exact value of $S_t$.
Numerical convergence tests described in ({\sl36}) show that for the high
($dx_{min} = 2.6\times10^5$~cm) and medium ($dx_{min} = 5.2\times10^5$~cm)
numerical resolutions the 3D structure of the turbulent flame was
resolved in sufficient detail to reproduce key dynamic properties
of the flame included in the subgrid model. Changing $L$ in Eq.~(2)
from $dx$ to $2dx$ has only a minor effect on the converged solution.
The results are self-consistent, reasonably accurate, and can be used
to analyze explosion scenarios for thermonuclear supernovae.

\bigskip
\noindent{\bf Thermonuclear deflagration in carbon-oxygen white dwarf 
of Chandrasekhar mass.}
The initial conditions for the simulations (see ({\sl36}) for
more details) were set up for a Chandrasekhar-mass WD in hydrostatic
equilibrium with the initial radius $R_{WD}=2\times 10^8 cm$, the initial
central density $\rho_c= 2\times 10^9$~g/cm$^3$, the uniform initial
temperature $T=10^5$~K, and the uniform initial composition with equal
mass fractions of $^{12}$C and $^{16}$O nuclei.  The 3D computational
mesh extended from the WD center $x=y=z=0$ to $x=y=z=2.6R_{WD}$. Thus,
we model one octant of the WD assuming mirror symmetry along the $x=0$,
$y=0$ and $z=0$ planes.  The burning was initiated at the center of WD
by filling a small spherical region at $r<0.015R_{WD}$ with hot reaction
products without disturbing the hydrostatic equilibrium.

The development of the explosion for the high-resolution case is shown
in Fig.~1 by a series of 3D snapshots of the flame and WD surfaces 
(see also Movies S1 and S2).
At the beginning, the initially spherical flame ignited in the center
of WD propagates outwards with the normal laminar flame speed $S_l$.
As it moves away from the center, the gravitational acceleration $g$ and
the Atwood number $A$ increase. This increases the amplitude and rate of
development of the RT instability controlled by $g$ and $A$. Due to the
RT instability, small perturbations of the flame surface grow and form
a few plumes that have characteristic mushroom shapes.  The turbulent
flame speed $S_t$ also increases with $g$ and $A$ according to Eq.(2), and
eventually dominates $S_l$ in Eq.(1). The flame plumes continue to grow,
due partially to the flame propagation and partially to gravitational
forces that cause the hot, burnt, low-density material inside the plumes to
rise towards the WD surface. The same gravitational forces also pull the
cold, high-density unburnt material between the plumes down towards the
center. The resulting shear flows along the flame surface are unstable
(Kelvin-Helmholtz (KH) instability) and quickly develop vortices. These
vortices further distort the flame surface, and also contribute their
energy into the turbulent cascade that creates turbulent
motions at smaller scales, down to a few $dx$.

When the original flame plumes grow large enough, secondary RT
instabilities develop on their surface, thus producing the next
level of ``mushrooms'' that also grow and may become subject to the
RT instability at a smaller scale, etc. These smaller gravity-induced
mushrooms interact with the turbulence created by the previous generation
of larger flame plumes, and also produce some turbulence themselves
through the KH instability.  The resulting complicated turbulent flame
surface is shown in Fig.~1.

As the turbulent flame develops, the energy released by the thermonuclear
burning causes the WD to expand. The expansion accelerates and becomes
nonuniform as the rising plumes approach the star surface.  We continued
the simulations until the star surface reached the computational
domain boundary.  By that time, the radius of the expanding star
increased by about a factor of 2.6, the outer layers accelerated to about
$1.2\times10^9$~cm/s, and the density of unburnt material near the star
center decreased to about $5\times10^7$~g/cm$^3$.  The area around the
center still contains a significant amount of unburnt material that
sinks at $10^8$~cm/s towards the center between large flame plumes.
The velocity of the large flame plumes is essentially zero relative
to the expanding matter, that is, the plumes have practically stopped
rising. This effect of freezing of the RT-instability on large scales
due to expansion is also related to freezing of large-scale turbulence
({\sl37}), and contributes to the burning-rate decrease after 1.5~s.

\bigskip
\noindent{\bf Effects of initial composition and background turbulence.}
Before we compare the computed supernova explosion with astronomical
observations, we need to know the extent to which the results are affected
by uncertainties in initial model parameters. We examined two properties
of the WD that are not well defined and are expected to have the most
pronounced effect on the explosion, the initial composition and the
level of background turbulence. 

In the base case described above, we assumed a uniform composition
with the carbon mass fraction $X_C=0.5$.  A WD can also have a core
partially depleted of carbon ({\sl42}).  To take this into account,
we calculated the explosion for $X_C=0.25$ in the core and $X_C=0.5$ in
the outer layers. The core with the radius $0.25R_{WD}$ contained about
25\% of the total mass of the WD. The lower C concentration resulted
in a slightly slower explosion as shown in Fig.~2 by the total energy
$E_{tot}$, the kinetic energy $E_k$, the released nuclear energy $E_n$,
and the burnt mass fraction $f_b$ as functions of time.  Compared to the
base case, it takes about 0.2 seconds longer for the slower explosion
to release the same amount of energy. There is no significant difference
in the turbulent flame structure.

The background turbulence is created by intense convective flows that
appear when the accreting WD approaches the Chandrasekhar limit and
quickly contracts just before ignition. A flame propagating through
intense turbulent flow can be accelerated to the velocity of
turbulent motions, $V_t$. Two-dimensional simulations ({\sl42}) show that
$V_t$ estimated as a differential velocity between adjoining eddies with
sizes $\sim10^7$~cm can reach $2.4\times10^7$~cm/s in central parts
of a WD. To model this effect, we replaced Eq.(1) by
$$
S = \max ( S_l, S_t, V_t )                                   \eqno(3)
$$
\noindent
and assumed that $V_t$ has a constant value $V_t^0$ for $r < 0.33R_{WD}$,
and linearly decreases to 0 as $r$ approaches $R_{WD}$.  The simulations
were performed for $V_t^0=3\times10^7$~cm/s.  The results (Fig.~2) show
that the background turbulence significantly increases the energy-release
rate and the expansion at the initial stage of the explosion. During
the first 0.45 s, the explosion with the background turbulence releases
the same amount of energy as the base case during 0.8 s. The material
burns faster because the laminar flame speed ($\sim10^7$~cm/s), which
played an important role at the initial stage for the base case, was
replaced by a higher value of $V_t$ according to Eq.(3).  The initial
flame speed, however, also has a long-lasting effect because the initial
stage of explosion is critical for the development of the turbulent flame
surface. A flame that initially propagates with a higher speed competes
with the RT instability more efficiently and develops fewer wrinkles on
resolved scales.  This results in a smaller surface area of the flame.
As the flame propagates outwards, $S_t$ increases for both cases. It
becomes higher than $S_l$ for the base case, and higher than $V_t$ for
the case with the background turbulence. $S_t$ is approximately the same
for both cases, but the explosion with the background turbulence begins
to slow down when the burnt mass fraction $f_b$ reaches 0.15 because of
a smaller flame surface area, and eventually releases much less energy
than the base case.

If we increase $V_t^0$ to the unrealistically high value $10^8$~cm/s,
the background turbulence dominates even far from the center and the
RT instability does not develop at all. The flame surface is almost
spherical until the end of the simulation, and there is no unburnt
material below the flame surface.  The energy release and expansion are
fast (Fig.~2) until the flame reaches low-density layers where the C
burning almost stops.  The final values of $E_{tot}$, $E_k$, $E_n$, and
$f_b$ are higher than for $V_t^0=3\times10^7$~cm/s, but lower than for
the base case in which the highly convoluted turbulent flame continues
to burn high-density material in central parts of the WD.

The uncertainty in the maximum energy released by the explosion can
be estimated as the difference between the base case and the case
with $V_t^0=3\times10^7$~cm/s (Fig.~2). Even though the burning
continues for the base case, it begins to slow down and will stop
when the density of the unburnt material near the center drops below
$\simeq10^6$~g/cm$^3$. We expect that by that time, the burnt mass
fraction will increase to 0.6-0.7, which is 1.7-2 times higher than for
the case with the background turbulence. The final released energies
for these two cases will also differ by a factor of 1.7-2 ($E_n \simeq
0.87\times10^{51}$~ergs for $V_t^0=3\times10^7$~cm/s, and $E_n \simeq
(1.5-1.7)\times10^{51}$~ergs for the base case).

\bigskip
\noindent{\bf Comparison of simulations and observations.}
Now, after ensuring numerical convergence and examining the sensitivity
to initial conditions, we can try to compare results of simulations
to astronomical observations.  One important parameter that can be
directly compared to observations is the kinetic energy of the expanding
material. This energy can be calculated from expansion velocities measured
using the Doppler shift of spectral lines. The typical kinetic energy
for SN~Ia is $(1-1.5)\times10^{51}$~ergs ({\sl6}). In simulations, this
energy corresponds to $E_k$ at infinity, when all the nuclear energy
released transforms into kinetic energy of the expanding material and
the work against gravity: $E_k = E_n + E^0_{tot}$. The negative initial
value $E^0_{tot}=-0.5\times10^{51}$ represents the binding energy of
the star.  For our base case, the estimated final value $E_k = E_n -
0.5\times10^{51} \simeq (1.0-1.2)\times10^{51}$~ergs is at the lower end
of the typical range.  The case with the realistic level of background
turbulence, $V_t^0=3\times10^7$~cm/s, produces a weak explosion, with
final $E_k \simeq 0.37\times10^{51}$~ergs.  The explosion is also not
strong enough ($E_k = 0.6\times10^{51}$~ergs) for the high level of
background turbulence, $V_t^0=10^8$~cm/s.  For all the cases shown in
Fig.~2, the final value of $E_{tot}$ is positive, that is, the explosion
releases enough energy to unbind the star.  The base-case 3D explosion
is much more energetic than explosions resulted from two-dimensional
simulations ({\sl43,26,34}) because the flame does not develop enough surface
area in two dimensions as discussed in ({\sl44}).

The key feature of the simulations is the highly convoluted turbulent
flame surface that allows extensive interpenetration of burnt and unburnt
materials. The angle-averaged burnt mass fraction at 1.9~s is less that 1
for any distance $r$ from the WD center (Fig.~3). In particular, $80-90\%$
of the material near the WD center is unburnt.  This material continues
to burn, but it will not burn out completely as long as convective flows
supply fresh unburnt material from outer layers. As the WD continues
to expand, the density decreases. The deflagration begins to produce
intermediate-mass elements when the density of unburnt material becomes
lower than $\simeq5\times10^7$~g/cm$^3$, and the burning stops when the
density drops below $\simeq10^6$~g/cm$^3$. This means that the
final ejecta produced by the 3D deflagration model will contain the
unburnt material and intermediate-mass elements at any distance from
the center.

The unburnt carbon and oxygen that remain between the flame plumes and
intermedi\-ate-mass elements that form at low densities at different
$r$ should produce spectral signatures in a wide range of expansion
velocities, including low velocities close to zero.
Analyses of SN~Ia spectra, however, imply C ({\sl45-47}) and O ({\sl48})
only at high velocities, as would be produced by the acceleration of
expanding outer layers.  For intermediate-mass elements, minimum observed
velocities are lower ({\sl2}), but large enough ($\sim10,000$~km/s
for Si) to rule out the presence of these elements near the WD center.
In the simulations, the low-velocity unburnt material was not present only
for the case with an unrealistically high level of background turbulence
that produced a spherical flame and resulted in a weak explosion.

The dynamics of the 3D deflagration described here is in agreement with
the preliminary simulations ({\sl33}) performed with the same model. These
results were independently confirmed by Reinecke et al. ({\sl34,35})
who used different subgrid model, nuclear kinetics scheme and initial
conditions, and obtained similar highly convoluted flames with unburnt
material near the WD center.

We thus conclude that the deflagration model of a SN~Ia explosion is
incomplete. The most natural solution to this problem that would make
the results consistent with observations would be to assume that the
turbulent flame triggers a detonation.  A thermonuclear detonation
wave could propagate through the WD with velocities $\sim10^9$~cm/s
({\sl49,50}) and would quickly burn all the material near the center
leaving only the low-density outer layers unburnt. For the density below
$5\times10^7$~g/cm$^3$, a detonation would produce intermediate-mass
elements ({\sl25}) that are observed in spectra of SNe~Ia.  A detonation
would also partially smooth out composition inhomogeneities that are
predicted by the deflagration model and may be incompatible with
observations ({\sl51}). Remaining asymmetries may account for a weak
polarization recently detected in SN~Ia spectra ({\sl52,53}).

One-dimensional ({\sl25,28-32}) and two-dimensional ({\sl26,27})
delayed-detonation models were the most successful in explaining
observable characteristics of SNe~Ia. These models, however, use the time
for detonation initiation as a free parameter because the DDT problem is
intrinsically 3D and still unsolved.  A large-scale 3D model also cannot
reproduce DDT phenomena that involve physical processes occurring on small
unresolved scales.  One approach to solving this problem is to study, in
much more detail, the types of reacting flows created by 3D deflagrations
and look for situations that create the right types of ``hot spots''
that we know ({\sl54}) are the sources of detonation initiation.

\vfill\eject

{\bf References and Notes}
\bigskip

\item {1.}          
J.~C.~Wheeler, R.~P.~Harkness, {\sl Rep. Prog. Phys.} {\bf 53}, 1467 (1990).
 
\item {2.}       
A.~V.~Filippenko, {\sl Annu. Rev. Astron. Astrophys.} {\bf 35}, 309 (1997)

\item {3.}       
J.~C.~Wheeler, {\sl Am. J. Phys.}, in press (2002)

\item {4.}
S.~E.~Woosley, T.~A.~Weaver, {\sl Annu. Rev. Astron. Astrophys.} {\bf 24},
205 (1986)

\item {5.}
D.~Branch, A.~M.~Khokhlov, {\sl Phys. Rep.} {\bf 256}, 53 (1995)

\item {6.}
J.~C.~Wheeler, R.~P.~Harkness, A.~M.~Khokhlov, P.~A.~H\"oflich,
{\sl Phys. Rep.} {\bf 256}, 211 (1995)

\item {7.}
K.~Nomoto, K.~Iwamoto, N.~Kishimoto, {\sl Science} {\bf 276}, 1378 (1997)

\item {8.}
D.~Branch,  {\sl Annu. Rev. Astron. Astrophys.} {\bf 36}, 17 (1998)

\item {9.}
W.~Hillebrandt, J.~C.~Niemeyer, {\sl Annu. Rev. Astron. Astrophys.}
{\bf 38}, 191 (2000)

\item {10.}
A.~Burrows, {\sl Nature} {\bf 403}, 727 (2000)

\item {11.}
F.~Hoyle, W.~A.~Fowler, {\sl Astrophys. J.} {\bf 132}, 565 (1960).

\item {12.}
M.~M.~Phillips, {\sl Astrophys. J.} {\bf 413}, L105 (1993) 

\item {13.}
M.~Hamuy {\sl et al}., {\sl Astron. J.} {\bf 109}, 1 (1995) 

\item {14.}
A.~G.~Riess, W.~H.~Press, R.~P.~Kirshner, {\sl Astrophys. J.} {\bf 438}, L17 (1995)

\item {15.}
S.~Perlmutter {\sl et al}., {\sl Nature} {\bf 391}, 51 (1998)

\item {16.}
S.~Perlmutter {\sl et al}., {\sl Astrophys. J.} {\bf 517}, 565 (1999)

\item {17.}
B.~P.~Schmidt {\sl et al}., {\sl Astrophys. J.} {\bf 507}, 46 (1998)

\item {18.}
A.~G.~Riess {\sl et al}., {\sl Astron. J.} {\bf 116}, 1009 (1998)

\item {19.}
A.~G.~Riess {\sl et al}., {\sl Astrophys. J.} {\bf 560}, 49 (2001)

\item {20.}       
P.~A.~H\"oflich {\sl et al}., {\sl Astrophys. J.} {\bf 472}, L81 (1996)

\item {21.}
W.~D.~Arnett, {\sl Astrophys. Space Sci.} {\bf 5}, 180 (1969)

\item {22.}
C.~J.~Hansen, J.~C.~Wheeler, {\sl Astrophys. Space Sci.} {\bf 3}, 464 (1969)

\item {23.}
K.~Nomoto, D.~Sugimoto, S.~Neo, {\sl Astrophys. Space Sci.} {\bf 39}, L37 (1976)

\item {24.}
K.~Nomoto, F.-K.~Thielemann, K.~Yokoi, {\sl Astrophys. J.} {\bf 286}, 644 (1984)

\item {25.} 
A.~M.~Khokhlov, {\sl Astron. Astrophys.} {\bf 245}, 114 (1991)

\item {26.}
D.~Arnett, E.~Livne, {\sl Astrophys. J.} {\bf 427}, 315 (1994)

\item {27.}
D.~Arnett, E.~Livne, {\sl Astrophys. J.} {\bf 427}, 330 (1994)

\item {28.}
H.~Yamaoka, K.~Nomoto, T.~Shigeyama, F.-K.~Thielemann,
{\sl Astrophys. J.} {\bf 393}, L55 (1992)

\item {29.}
A.~M.~Khokhlov, E.~M\"uller, P.~A.~H\"oflich, {\sl Astron. Astrophys.}
{\bf 270}, 223 (1993)

\item {30.}
P.~A.~H\"oflich, {\sl Astrophys. J.} {\bf 443}, 89 (1995)

\item {31.}
P.~A.~H\"oflich, A.~M.~Khokhlov, J.~C.~Wheeler,
{\sl Astrophys. J.} {\bf 444}, 831 (1995)

\item {32.}
P.~A.~H\"oflich, A.~M.~Khokhlov, {\sl Astrophys. J.} {\bf 457}, 500 (1996)

\item {33.}
A.~M.~Khokhlov, http://www.arxiv.org/abs/astro-ph/0008463 (2000)

\item {34.}
M.~Reinecke, W.~Hillebrandt, J.~C.~Niemeyer, {\sl Astron. Astrophys.}
{\bf 386}, 936 (2002)

\item {35.}
M.~Reinecke, W.~Hillebrandt, J.~C.~Niemeyer, {\sl Astron. Astrophys.},
{\bf 391}, 1167 (2002)

\item {36.}
Materials and methods are available as supporting material on {\sl Science} Online.

\item {37.}
A.~M.~Khokhlov, {\sl Astrophys. J.} {\bf 449}, 695 (1995)

\item {38.}
A.~M.~Khokhlov, {\sl J. Comput. Phys.} {\bf 143}, 519 (1998)

\item {39.}
F.~X.~Timmes, S.~E.~Woosley, {\sl Astrophys. J.} {\bf 396}, 649 (1992)

\item {40.}
A.~M.~Khokhlov, E.~S.~Oran, J.~C.~Wheeler, {\sl Astrophys. J.}
{\bf 478}, 678 (1997)

\item {41.}
A.~M.~Khokhlov, E.~S.~Oran, J.~C.~Wheeler, {\sl Combust. Flame}
{\bf 105}, 28 (1996)

\item {42.}
P.~A.~H\"oflich, J.~Stein, {\sl Astrophys. J.} {\bf 568}, 779 (2002)

\item {43.}
E.~Livne, {\sl Astrophys. J.} {\bf 406}, L17 (1993)

\item {44.}
A.~M.~Khokhlov, {\sl Astrophys. J.} {\bf 424}, L115 (1994)

\item {45.}
D.~J.~Jeffery {\sl et al}., {\sl Astrophys. J.} {\bf 397}, 304 (1992)

\item {46.}
A.~Fisher, D.~Branch, P.~Nugent, E.~Baron, {\sl Astrophys. J.} {\bf 481}, L89 (1997)

\item {47.}
P.~A.~Mazzali, {\sl Mon. Not. R. Astron. Soc. } {\bf 321}, 341 (2001)

\item {48.}
R.~P.~Kirshner {\sl et al}., {\sl Astrophys. J.} {\bf 415}, 589 (1993)

\item {49.}
A.~M.~Khokhlov, {\sl Mon. Not. R. Astron. Soc. } {\bf 239}, 785 (1989)

\item {50.}
V.~N.~Gamezo, J.~C.~Wheeler, ~A.~M.~Khokhlov, E.~S.~Oran,
{\sl Astrophys. J.} {\bf 512}, 827 (1999)

\item {51.}
R.~C.~Thomas, D.~Kasen, D.~Branch, E.~Baron, {\sl Astrophys. J.} 
{\bf 567}, 1037 (2002)

\item {52.}
L.~Wang, J.~C.~Wheeler, P.~A.~H\"oflich, 
{\sl Astrophys. J.} {\bf 476}, L27 (1997)

\item {53.}
D.~A.~Howell, P.~A.~H\"oflich, L.~Wang, J.~C.~Wheeler,
{\sl Astrophys. J.} {\bf 556}, 302 (2001)

\item {54.}
A.~M.~Khokhlov, E.~S.~Oran, {\sl Combust. Flame} {\bf 119}, 400 (1999)

\item {55.}
This work was supported in part by the NASA ATP program
(NRA-99-01-ATP-130) and by the Naval Research Laboratory (NRL) through
the Office of Naval Research. Computing facilities were provided by
the DOD HPCMP program. We would like to thank David Branch, Peter
H\"oflich, Eli Livne, Martin Reinecke, Lifan Wang, and Craig Wheeler
for helpful discussions, and the referees for useful comments.

\bigskip
\noindent
{\bf Supporting Online Material}          

\noindent
www.sciencemag.org                          

\noindent
Materials and Methods                    

\noindent
Figs. S1, S2                             

\noindent
Table S1                                 

\noindent
Movies S1, S2

\vfill

\input epsf.sty

\bigskip
\centerline{\bf FIGURES }
\bigskip

\noindent
{\bf Fig.~1.}
Development of thermonuclear deflagration in carbon-oxygen white
dwarf.  The gray surface shows the turbulent thermonuclear flame.  The
color scale shows the radial velocity of unburnt material scaled by
1000~km/s.  Distances are scaled by the computational domain size
$x_{max} = 5.35\times10^8$~cm.  Numbers in the frame corners are time in
seconds (left top), burnt mass fraction $f_b$ (right top), and flame
surface area scaled by $x_{max}^2$ (right bottom).
High resolution ($dx_{min} = 2.6\times10^5$~cm). $L=2dx$.

\eject


\dimen1=8.0truecm
\dimen2=0.00truecm
\dimen3=0.1truecm

\vbox{ \vskip -1.5truecm

\centerline{
\hbox{
\epsfysize=\dimen1
\epsfbox{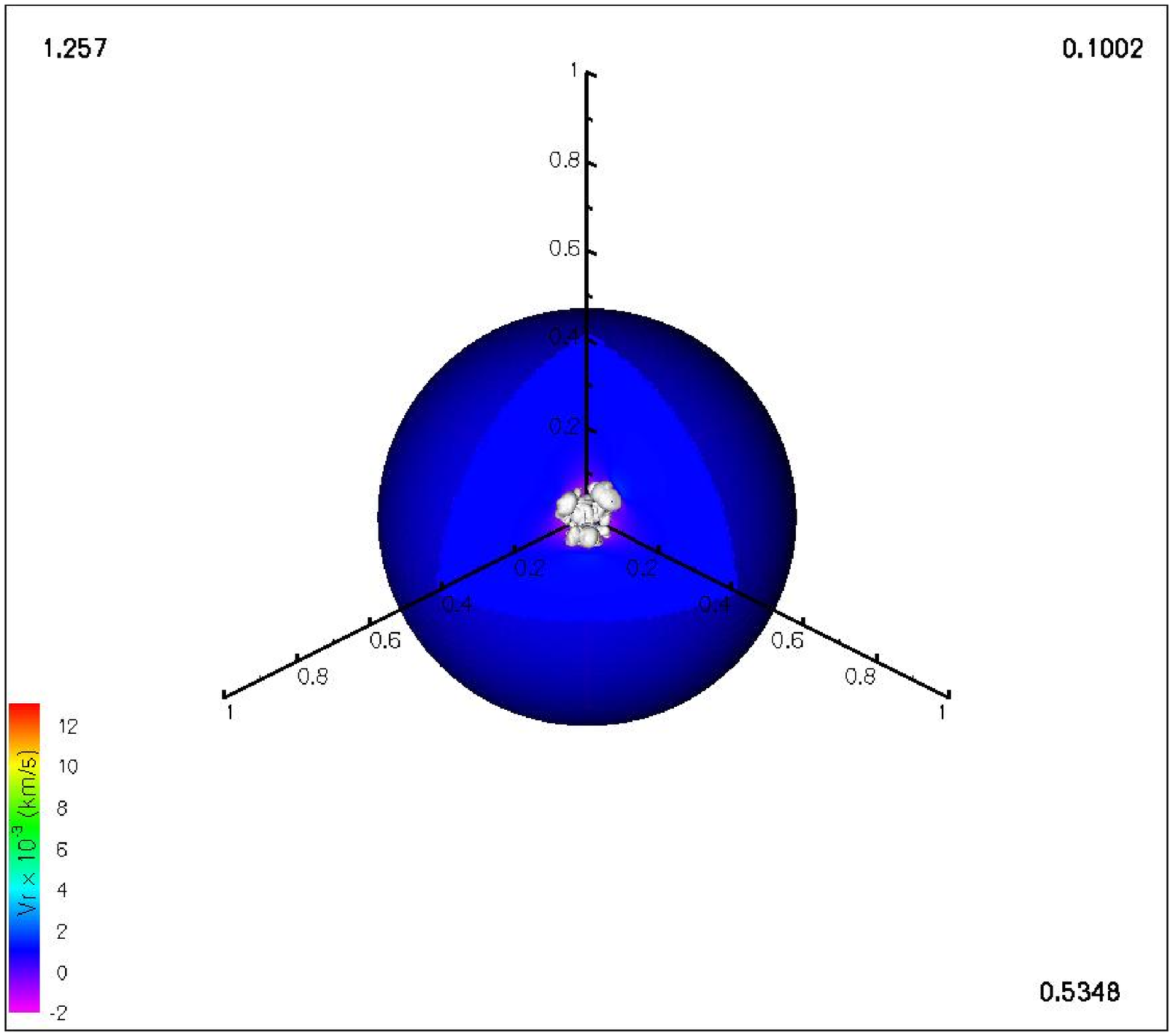}
\hskip \dimen2
\epsfysize=\dimen1
\epsfbox{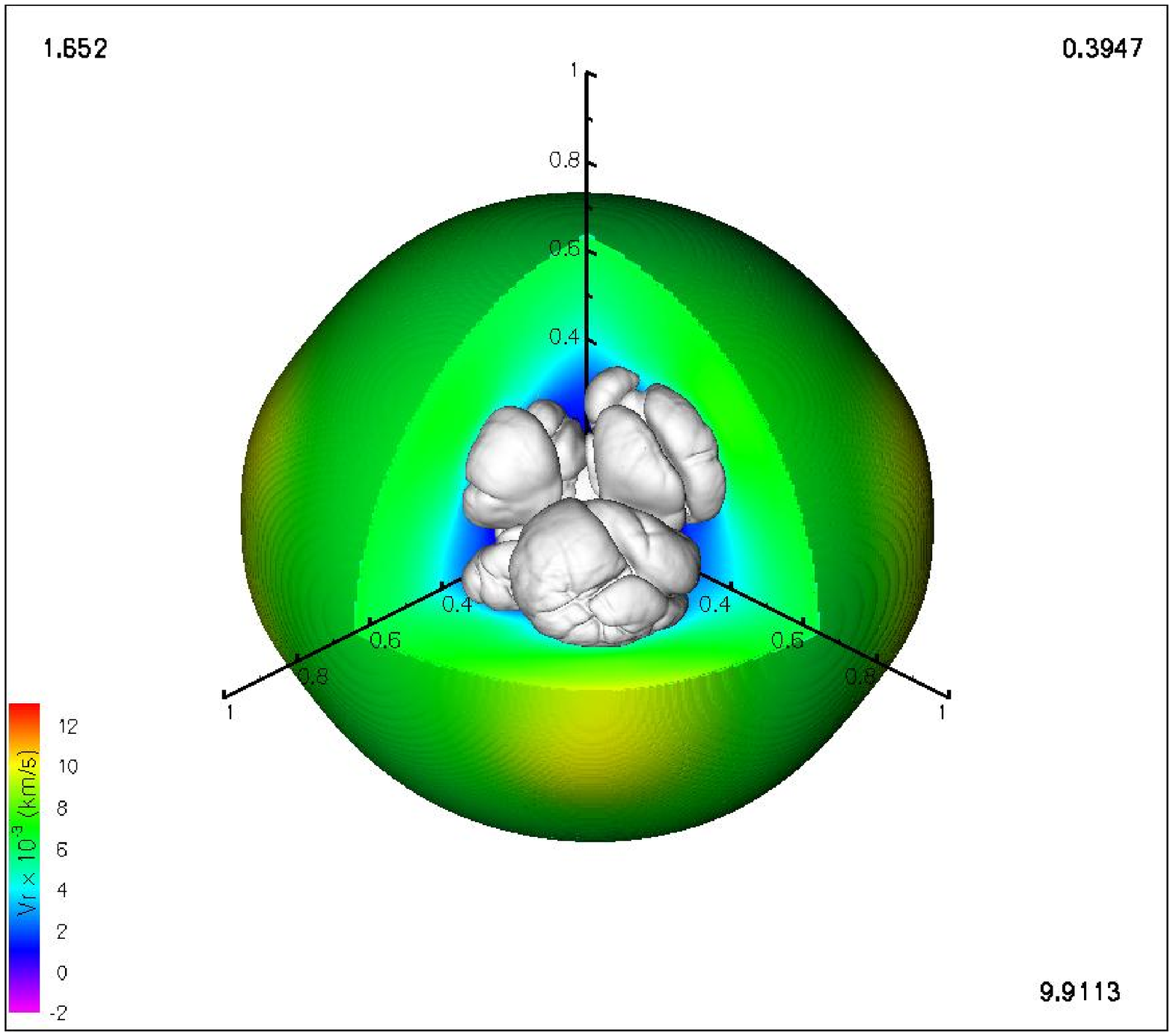}
} }
\vskip \dimen3

\centerline{
\hbox{
\epsfysize=\dimen1
\epsfbox{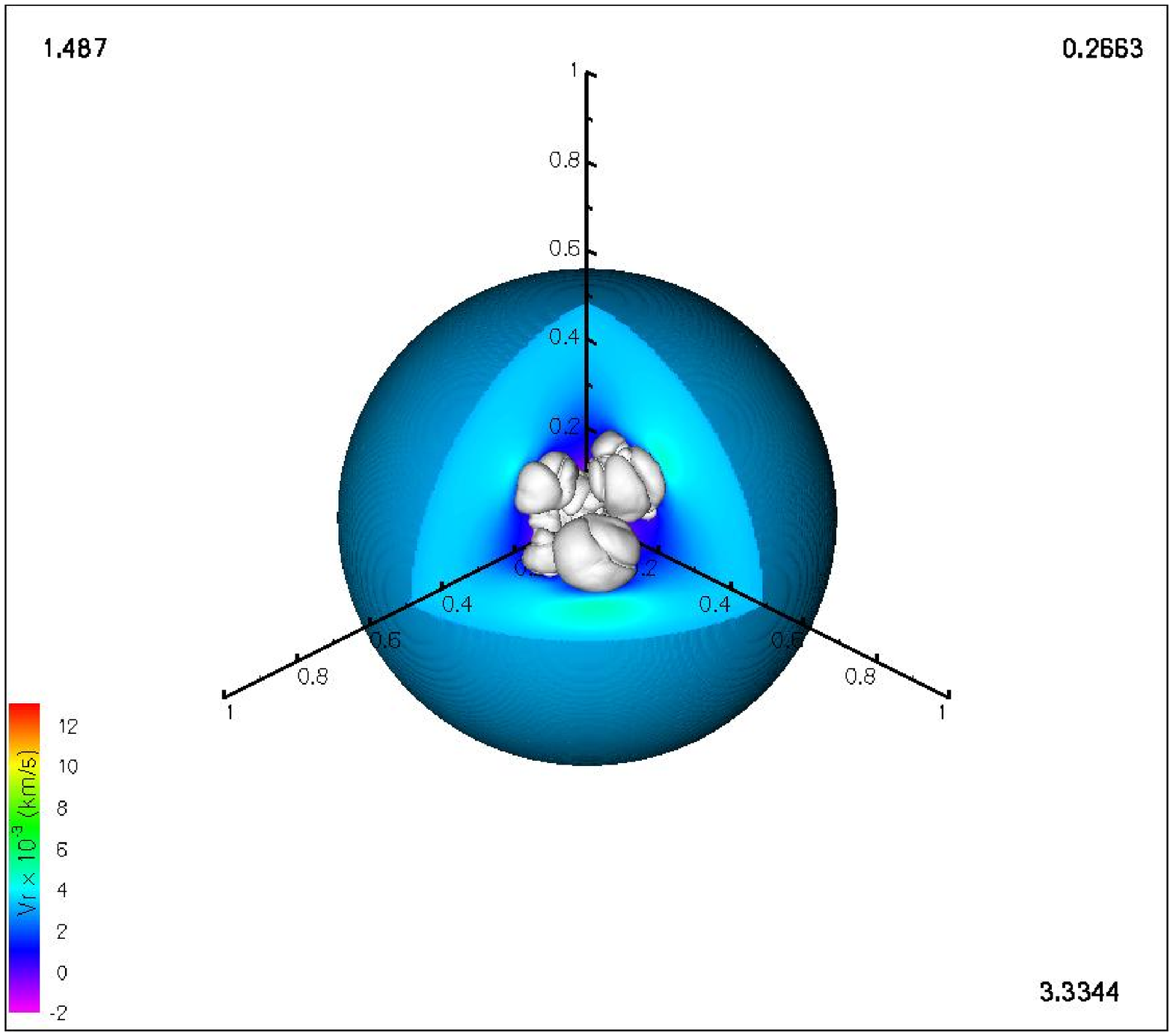}
\hskip \dimen2
\epsfysize=\dimen1
\epsfbox{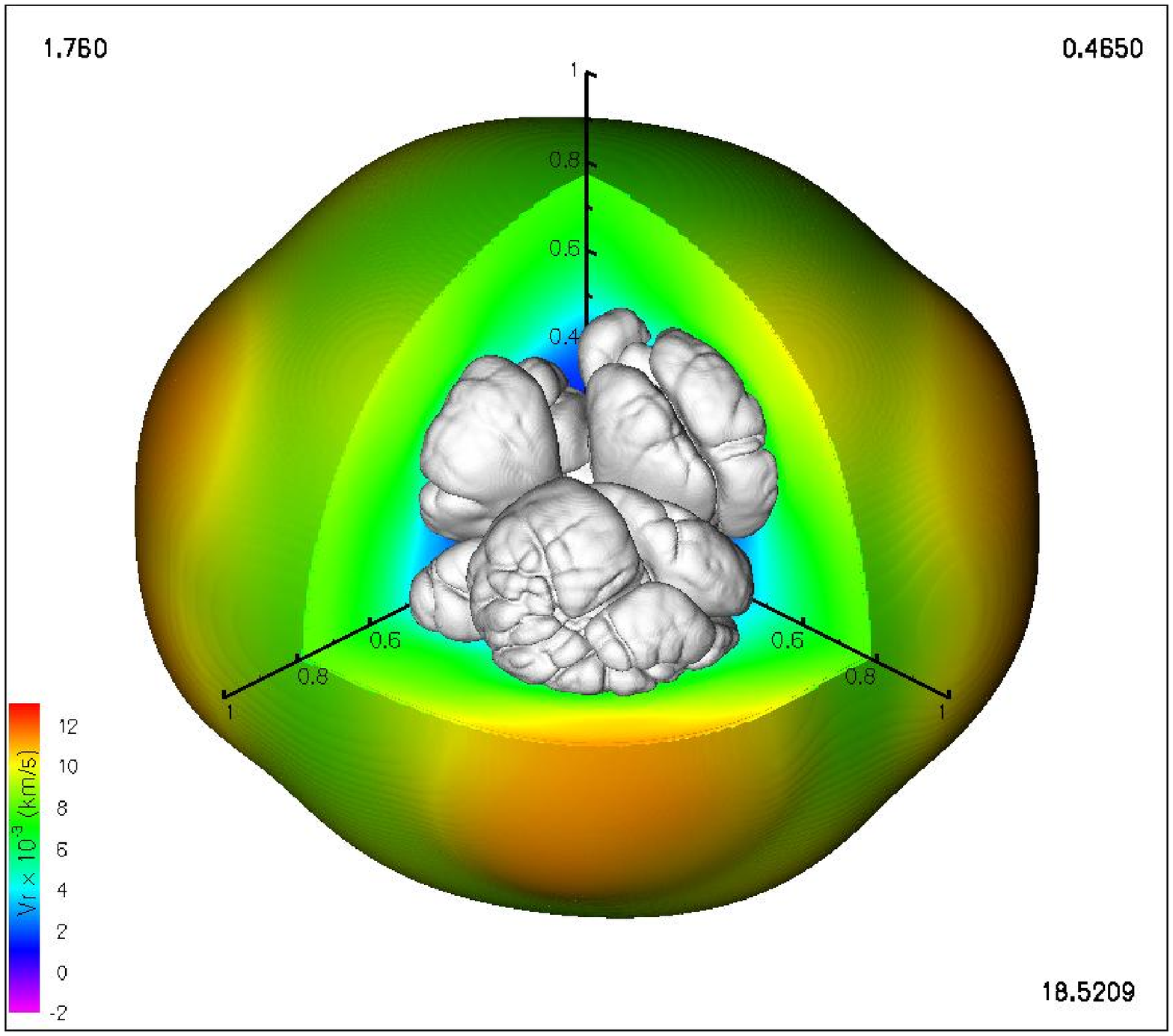}
} }
\vskip \dimen3

\centerline{
\hbox{
\epsfysize=\dimen1
\epsfbox{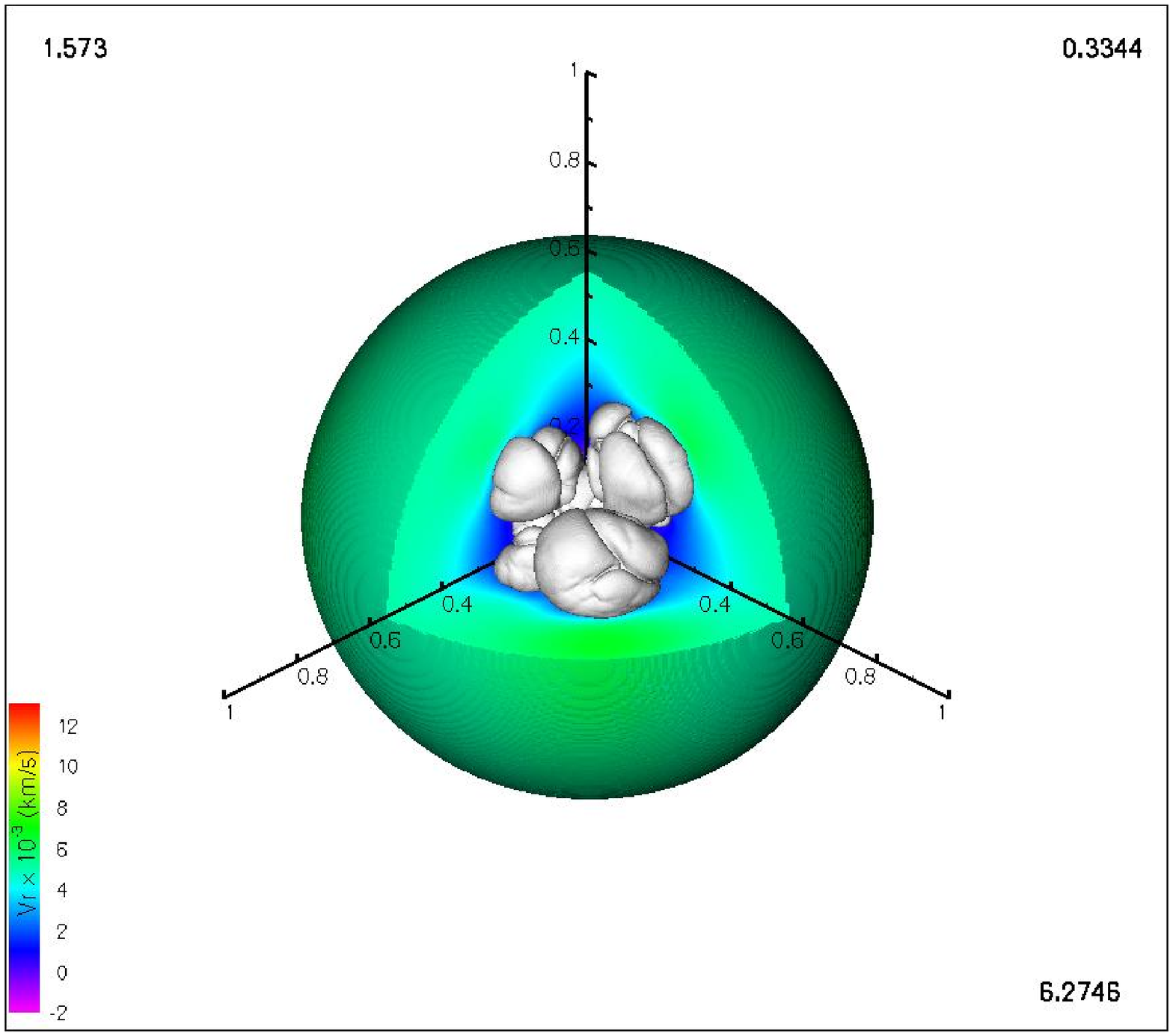}
\hskip \dimen2
\epsfysize=\dimen1
\epsfbox{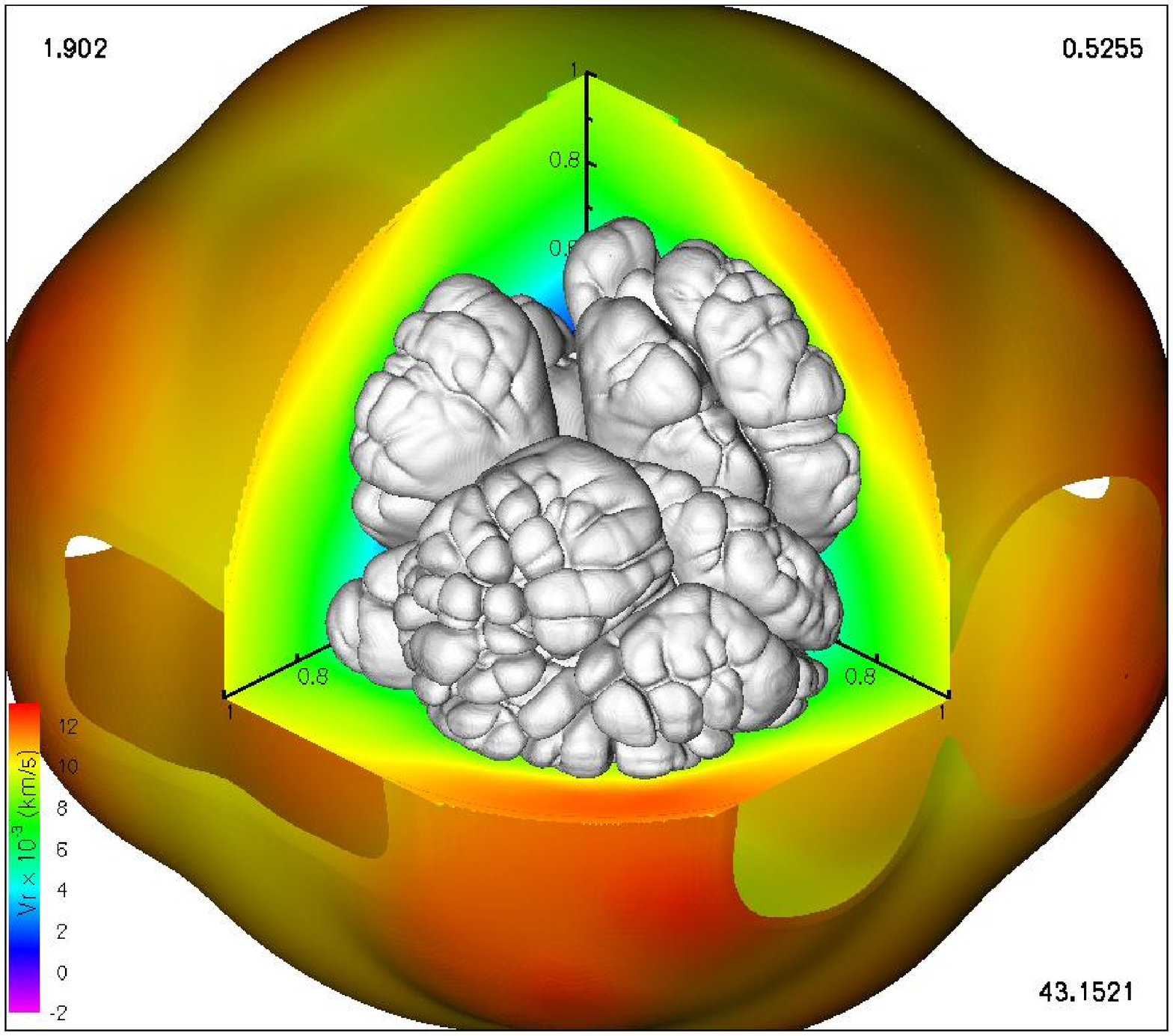}
} }
}
\vfill\eject


\centerline{
\hbox{
\epsfxsize=15truecm
\epsfbox{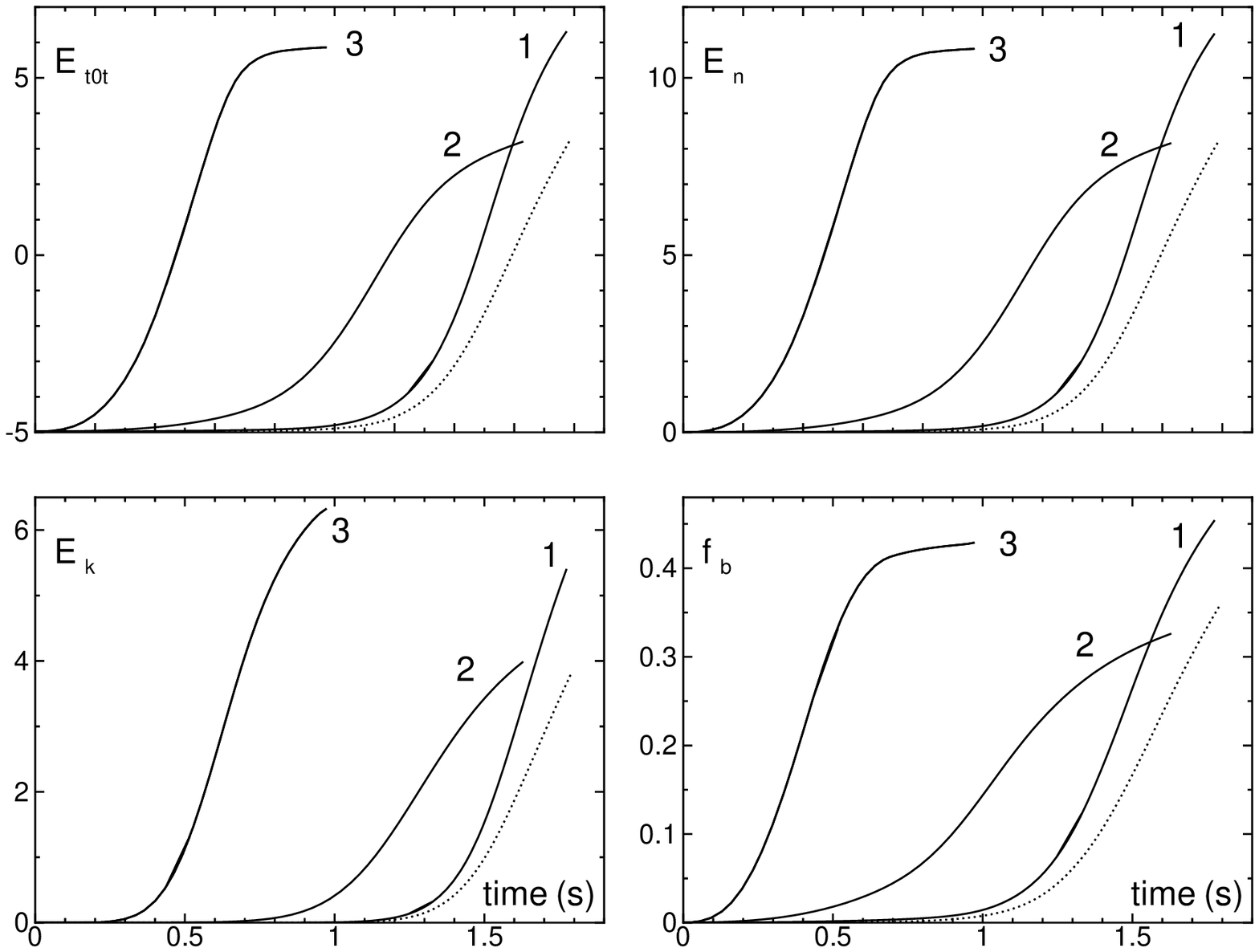}
} }

\noindent
{\bf Fig.~2.}
Total ($E_{tot}=E_k + E_t - E_g$), kinetic ($E_k$), released nuclear 
($E_n$) energies
and the burnt mass fraction $f_b$ as functions of time for the base
case without background turbulence (1), with background turbulence
$V_t^0=3\times10^7$~cm/s (2), and $V_t^0=10^8$~cm/s (3). Dotted lines
correspond to reduced carbon concentration $X_C=0.25$ in the core.
Energy units are $10^{50}$~ergs. Medium resolution 
($dx_{min} = 5.2\times10^5$~cm). $L=dx$.

\centerline{
\hbox{
\epsfxsize=9truecm
\epsfbox{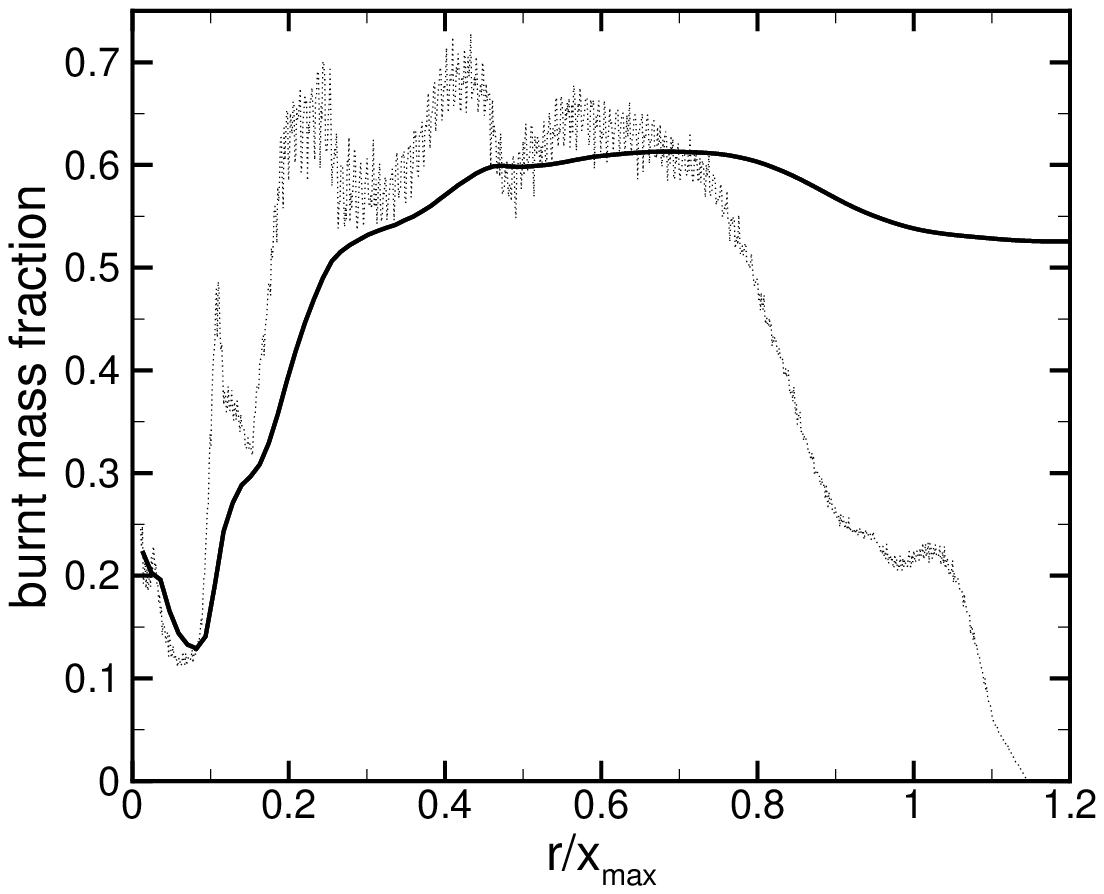}
} }

\noindent
{\bf Fig.~3.}
Angle-averaged local mass fraction of burnt material (dotted line) and
total burnt mass fraction inside a spherical region of radius $r$ (solid line)
as functions of distance $r$ from the WD center. The distance is scaled
by the computational domain size $x_{max} = 5.35\times10^8$~cm. Time 1.9~s
corresponds to the last frame of Fig.~1.

\vfill\eject

\noindent {\bf Supporting Online Material} 
\smallskip

\noindent
V. N. Gamezo, A. M. Khokhlov, E. S. Oran, A. Y. Chtchelkanova, R. O. Rosenberg.
Explosions of Thermonuclear Supernovae: Physical Implications of Full-Scale, 
Three-Dimensional Simulations.

\bigskip\bigskip
\centerline{\bf MATERIALS AND METHODS }

\bigskip
\noindent{\bf NRL Supernova Model }
\bigskip

\noindent{\sl Fluid Dynamics and Equation of State}
\bigskip

The numerical model is based on reactive Euler equations
$$
\eqalign{
{{\partial\rho}\over{\partial t}} &=
- \nabla\cdot\left(\rho {\bf U}\right)~,\cr
{{\partial\rho{\bf U}}\over{\partial t}} &=
- \nabla\cdot\left(\rho {\bf U}{\bf U}\right)
- \nabla P + \rho {\bf g} ~,\cr
{{\partial E}\over{\partial t}} &=
- \nabla\cdot\left({\bf U}\left( E+P\right)\right) +
  \rho {\bf U}\cdot{\bf g} + \rho \dot q~, \cr
}                
$$
where $\rho$, $E = E_i + \rho U^2/2$, $E_i$, ${\bf U}$, $\bf g$ and
$\dot q$ are the mass density, energy density, internal energy density,
velocity of matter,  gravitational acceleration, and nuclear energy
release rate per unit mass, respectively.  These equations describe
mass, momentum, and energy conservation laws for an inviscid fluid. The
thermodynamic properties of the fluid are defined by the equation of state
of degenerate matter, which is well described by basic theory and includes
contributions from ideal Fermi-Dirac electrons and positrons, equilibrium
Planck radiation, and ideal ions.  Pressure $P = P(\rho,E_i,Y_e,Y_i)$
and temperature $T=T(\rho,E_i,Y_e,Y_i)$ are determined by the equation of
state as functions of $\rho$, $E_i$, the electron mole fraction $Y_e$,
and the mean mole fraction of ions $Y_i$.  The relation between $P$
and $\rho$ is close to that in a polytropic gas with $\gamma$ varying
from 4/3 to 5/3.  This equation of state is valid for the thermodynamic
parameters and compositions expected in the computations, ranging from
relatively cold highly degenerate carbon-oxygen matter to partially
degenerate hot products of thermonuclear reactions.


\bigskip
\noindent{\sl Nuclear Kinetics and Flame Propagation}
\bigskip

Thermonuclear reactions involved in the thermonuclear burning of the
carbon-oxygen mixture ({\sl S1-S3}) can be separated into three consecutive
stages responsible for the energy release. First,  the $^{12}$C~+~$^{12}$C
reaction leads to the consumption  of $C$ and formation of mostly Ne,
Mg, protons, and $\alpha$-particles.  Then begins the nuclear statistical
quasi-equilibrium (NSQE) stage, during which $O$ burns out and Si-group
(intermediate mass) elements are formed.  Finally, Si-group elements
are converted into the Fe-group elements and the nuclear statistical
equilibrium (NSE) sets in. The reaction time scales associated with
these stages strongly depend on temperature and density and may differ
from one another by several orders of magnitude ({\sl S4-S8}).

The full nuclear reaction network includes hundreds of species that
participate in thousands of reactions.  Integration of this full
network is too time-consuming to be used in multidimensional numerical
models. Therefore, we used a simplified four-equation kinetic scheme
({\sl S9,S10}) that describes all major stages of carbon burning. This
kinetic scheme is coupled with a flame-capturing algorithm ({\sl S11,S10})
that ensures thermonuclear flame propagation with a prescribed speed
$S$. Flame capturing is needed because we cannot resolve in large-scale
simulations the physical thickness of a laminar flame that differs from
the white dwarf radius $R_{WD}$ by up to 12 orders of magnitude.

\bigskip
\noindent{\sl Numerical Method}
\bigskip

The fluid dynamic equations, coupled to the nuclear reaction
mechanism and the flame-capturing algorithm are integrated using an 
explicit, second-order, Godunov-type, adaptive-mesh-refinement code
({\sl S11,S12}). A Riemann solver is used to evaluate fluxes at
cell interfaces.  The computational mesh is comprised of cubic cells
of various sizes that are organized in a fully threaded tree (FTT)
({\sl S12}). The FTT-based parallel adaptive mesh refinement algorithm
dynamically adjusts cell sizes in accordance with changing physical
conditions in the vicinity of each cell.  Here, the mesh was refined
around shock waves, flame fronts, and in regions of steep gradients of
density, pressure, composition, and tangential velocity.  The cell size
$dx$ varies within predefined limits $dx_{min}$ and $dx_{max}$ in such
a way that neighboring cell sizes can be the same or differ by a factor
of two.  The code has been extensively tested and used in various combustion
problems involving shocks, flames, turbulence, and their interactions
({\sl S13,S14}, and references therein) and in astrophysics ({\sl S15}).

\bigskip
\noindent{\sl Initial Conditions}
\bigskip

The initial conditions for the simulations were set up for a
Chandrasekhar-mass white dwarf (WD) in hydrostatic equilibrium with the
initial radius $R_{WD}=2\times 10^8 cm$, the initial central density
$\rho_c= 2\times 10^9$~g/cm$^3$, the uniform initial temperature
$T=10^5$~K, and the uniform initial composition with equal mass
fractions of $^{12}$C and $^{16}$O nuclei.  Starting from the central
pressure $P(\rho_c)$, the equations of hydrostatic equilibrium, $dP/dr
= -GM\rho/r^2$ and $dM/dr=4\pi\rho r^2$, were integrated outward until
$P=0$ was reached (here G is the gravitational constant, and M is the
mass of the material inside a sphere of radius r).  The resulting WD
configuration was interpolated onto a 3D mesh extended from the WD
center $x=y=z=0$ to $x=y=z=2.6R_{WD}$. Thus, we model one octant of
the WD assuming mirror symmetry along the $x=0$, $y=0$ and $z=0$ planes.
The burning was initiated at the center of WD by filling a small spherical
region at $r<0.015R_{WD}$ with hot reaction products without disturbing
the hydrostatic equilibrium.

Because a Chandrasekhar-mass WD is close to a collapse threshold,  its
gravitational equilibrium is sensitive to the discretization errors
that appear when the spherical body is mapped into a Cartesian mesh.
To minimize these errors, the mesh is initially refined to the finest
level near the WD center ($r<0.4R_{WD}$). During the simulations, we
keep the mesh unchanged until the flame reaches the fine grid boundary.
Then the adaptive mesh refinement algorithm is turned on.

\bigskip
\noindent{\sl Subgrid Model for Flame Speed}
\bigskip

The speed of a laminar thermonuclear flame in a WD is defined by
reaction rates and transport properties of the material and governed
by the same laws that describe laminar flame structure in terrestrial
chemical systems ({\sl S16-S18}).  The only substantial difference is that
transport properties of degenerate matter are dominated by the electron
heat conduction at high densities, and by both electron and photon heat
conduction at low densities.  The steady-state laminar flame speed $S_l$
in carbon-oxygen degenerate matter is a known function of temperature,
density, and composition ({\sl S19,S20}), but the flame can be laminar
only near the center of a WD. Away from the center, the flame is turbulent
and propagates with a higher effective speed ({\sl S21,S11,S22,S23}).

In our simulations, the turbulent flame speed $S_t$ is defined by a
subgrid model that takes into account physical processes at scales smaller
than the computational cell size. The subgrid model assumes that burning
on small unresolved scales is driven by the gravity-induced Rayleigh-Taylor (RT)
instability.  This instability distorts the flame surface at multiple
scales and generates turbulent motions in the surrounding fluid. The
turbulent energy propagates from large to small scales, thus further
disturbing the flame surface on small scales.  A developed turbulent
flame is statistically steady-state and forms a dynamic hierarchical
self-similar 3D structure where the flame surface is distorted at multiple
scales.  The flame distortions increase the flame surface and, therefore,
the burning rate.  The larger the scale, the higher the burning rate
at this scale due to the increase of the flame surface resulting from
distortions at smaller scales.  The burning rate defines the turbulent
flame speed $S_t$ for any given length scale.  This turbulent flame
structure was analyzed in 3D numerical simulations ({\sl S11,S24}) performed
for the thermonuclear burning of carbon-oxygen degenerate matter in a
uniform gravitational field.  It was found that a turbulent flame in a
vertical column of width $L$ becomes quasi-steady-state and propagates
with the speed
$$
S_t \simeq 0.5 \sqrt{AgL}                         \eqno(S1)
$$
\noindent
independent of the laminar speed $S_l$, where $A=(\rho_0 - \rho_1)/(
\rho_0 + \rho_1)$ is the Atwood number, and $\rho_0$ and $\rho_1$ are  the
densities ahead and behind the flame front, respectively. We used this
result in the subgrid model assuming that at $L << R_{WD}$ burning can
be considered as locally steady state ({\sl S25,S26,S11}).  The driving scale
$L$ was set equal to the computational cell size $dx$.  The resulting flame
velocity $S$ used by the flame-capturing algorithm was calculated as
$$
S = \max ( S_l, S_t )                               \eqno(S2)
$$

\bigskip
\noindent{\bf Numerical Convergence}
\bigskip

Numerical convergence tests, which are critical for establishing the
validity of any numerical simulations, are performed to ensure that the
solution obtained is not significantly affected by the computational
cell size $dx$. In our case, numerical convergence is closely related
to the physics of the subgrid model that defines the turbulent flame
speed in Eq.(S1) and assumes that burning can be considered as locally
steady state. The steady-state assumption is not valid for length scales
affected by the spherical geometry and expansion. These scales need to
be explicitly resolved, but this will not necessarily ensure that the
solution is independent of $dx$.  The resolved flame properties should
also correspond to the flame properties built into the subgrid model.

The subgrid model is based on the two main properties of a turbulent flame
({\sl S11,S24}): self-similarity of the flame structure and self-regulation
of the flame speed.  Self-similarity means that the 3D distortions of
the flame surface at different scales are similar.  Self-regulation
means that changing the flame speed at small scales does not affect
the flame speed at larger scales. This occurs because a higher flame
speed at small scales causes small flame wrinkles to burn out, thus
decreasing the flame surface.  The resulting burning rate, defined as
a product of the flame speed at small scales and the flame surface,
does not change. This subgrid model makes it possible to reproduce the
correct flame propagation in numerical simulations while explicitly
resolving only the large-scale flame structure.  If the resolved flame
structure is self-similar and self-regulating, and behaves according to
the Eq.(S1), the subgrid model just extends this behavior to unresolved
small scales.  Shifting the boundary between resolved and unresolved
scales by changing the numerical resolution should not affect the
turbulent flame propagation.  The solution obtained should then be
independent of numerical resolution and on the exact value of $S_t$.

We performed a series of computations with three resolutions:  low
($dx_{min} = 10.5\times10^5$~cm), medium ($dx_{min} = 5.2\times10^5$~cm),
and high ($dx_{min} = 2.6\times10^5$~cm).  The maximum computational cell
size $dx_{max} = 41.8\times10^5 $~cm and the computational domain size
$x_{max} = 5.35\times10^8$~cm were kept constant.  The development of the
explosion for the high-resolution case is shown in Fig.~1 and described
in the main text.  The simulation with medium resolution showed a similar
pattern of development, though it produced fewer small-scale wrinkles on
the flame surface.  The low-resolution simulation showed only the largest
flame plumes.  The surface of these plumes remained smooth and did not
develop any significant instabilities. These three cases are compared
in Fig.~S1, which shows the total energy $E_{tot}$, the kinetic energy
$E_k$, the released nuclear energy $E_n$, and the burnt mass fraction
$f_b$ as functions of time for the developing explosions.  The curves for
high and medium resolutions practically coincide for $E_{tot}$, $E_n$,
and $f_b$.  The kinetic energy $E_k$ shows the largest difference between
the high-resolution and the medium-resolution cases, and this does not
exceed 10\%. Therefore, the solution is practically converged.

In order to test the self-regulating properties of the flame, we
increased $L$ in Eq.(S1) by a factor of 2 ($L=2dx$), and repeated
the three simulations with the same high, medium, and low numerical
resolutions. The key energies and $f_b$ for these simulations are compared
with previous three cases in Fig.~S2.  For the low-resolution case,
the self-regulation mechanism on resolved scales does not work, $S_t$
and key energies increase by the same factor $\sqrt2$.  For the medium
and high resolutions, the self-regulation appears on resolved scales
and reduces the maximum difference between the energies calculated for
different $S_t$ to 15\% and 6\%, respectively.  These tests show that the
numerical convergence was achieved in the high- and medium-resolution
simulations. The 3D structure of the turbulent flame was resolved in
sufficient detail to reproduce key dynamic properties of the flame
included in the subgrid model.  The results are self-consistent and
reasonably accurate, and can be used to analyze explosion scenarios for
thermonuclear supernovae.

\vfill\eject

\bigskip
\centerline{\bf SUPPORTING FIGURES }
\smallskip 

\input epsf.sty
\centerline{
\hbox{
\epsfxsize=17truecm
\epsfbox{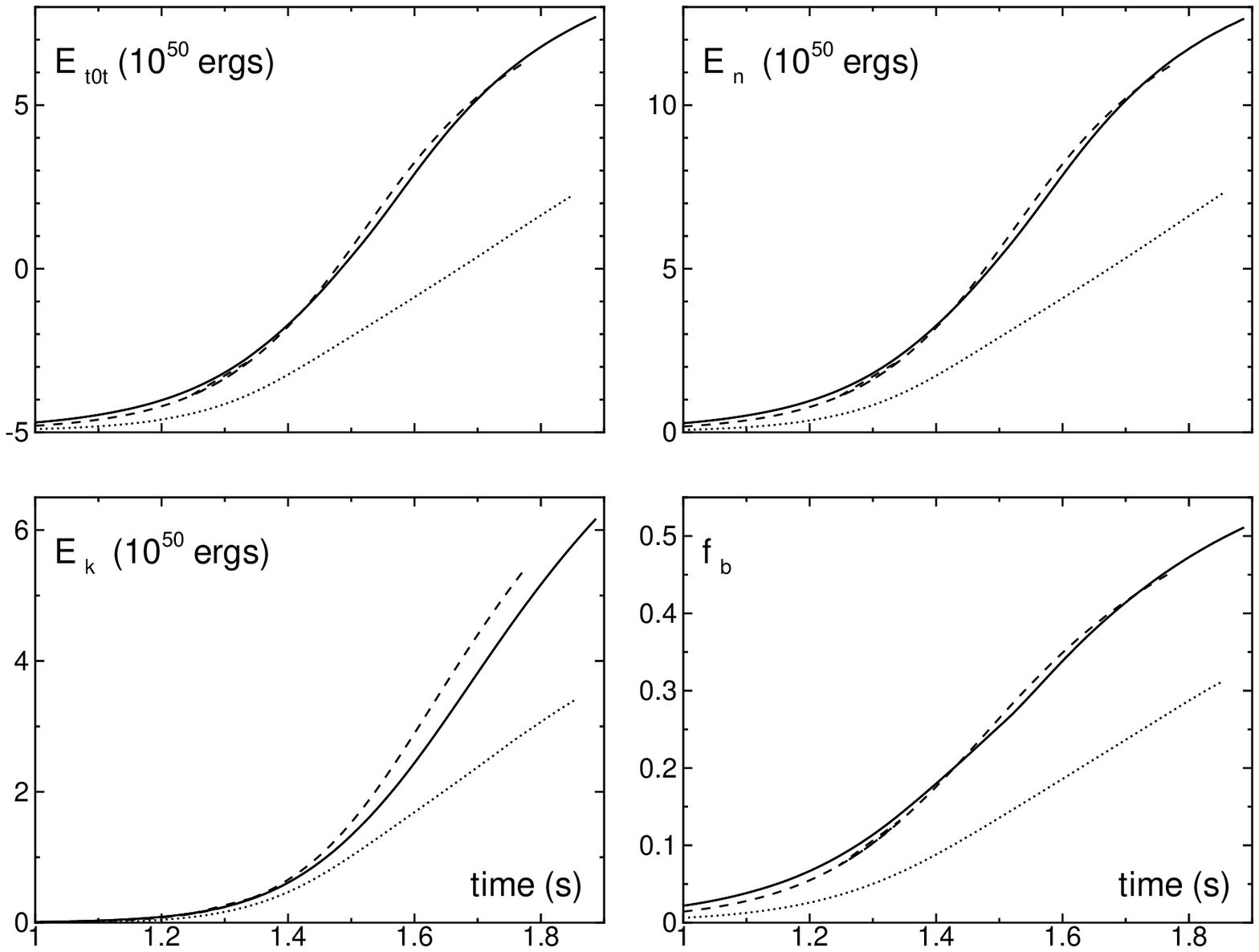}
} }
\vskip 1truecm

\noindent{\bf Fig.~S1.}
Total ($E_{tot}$), kinetic ($E_k$), released nuclear
($E_n$) energies and the burnt mass fraction $f_b$ as functions of
time. Solid, dashed, and dotted lines correspond to high, medium, and
low numerical resolution, respectively. $E_k$ is calculated from local
fluid velocities and densities as a total for all computational cells.
The conservation law $E_k + E_t - E_n -  E_g = const$ is satisfied with
accuracy better than 1\%.

\centerline{
\hbox{
\epsfxsize=10truecm
\epsfbox{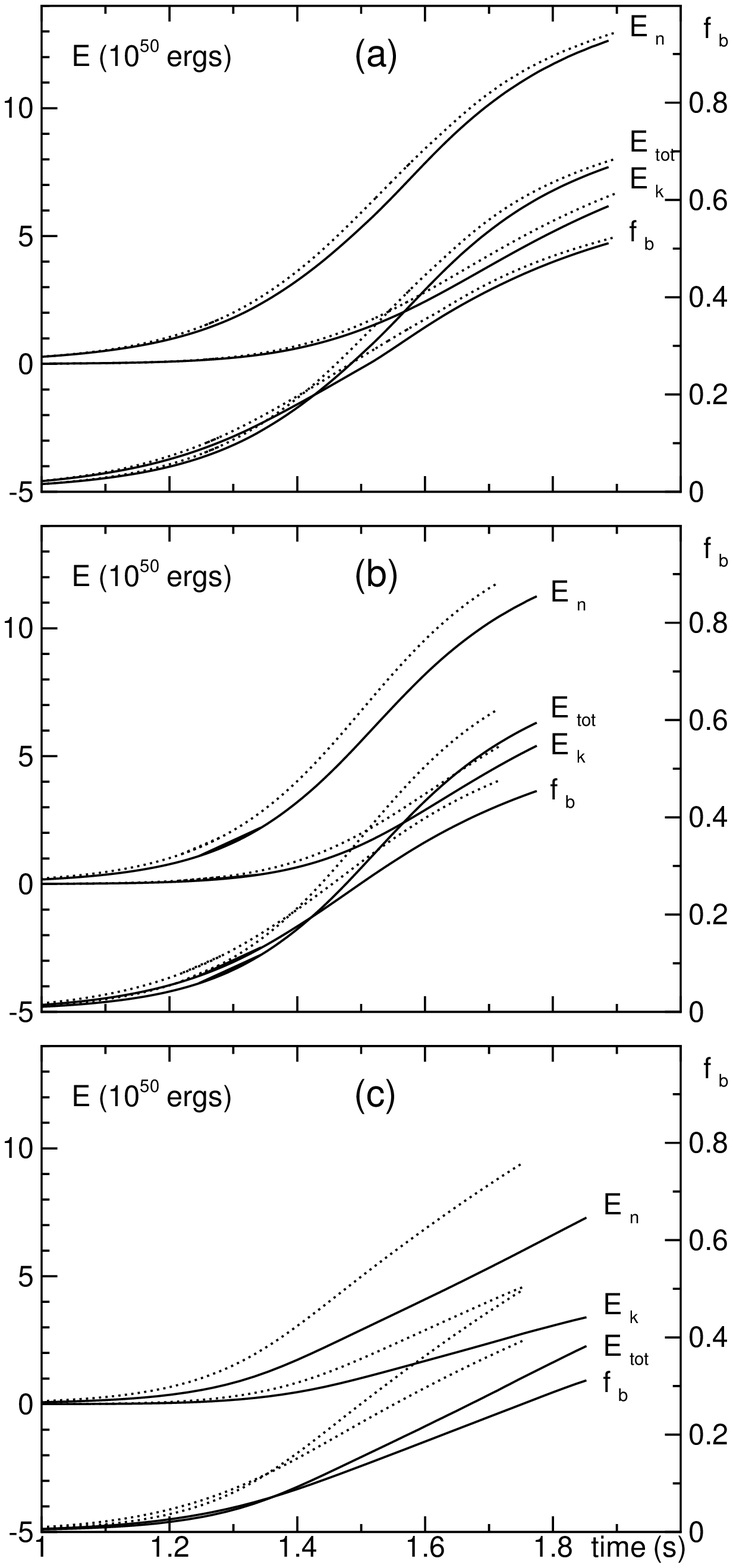}
\hskip 0.5truecm
\vbox{\hsize=7truecm
\tolerance=1000
\noindent{\bf Fig.~S2.}
Total ($E_{tot}$), kinetic ($E_k$), released nuclear ($E_n$) energies
and the burnt mass fraction $f_b$ as functions of time for high (a),
medium (b) and low (c) numerical resolutions, $L=dx$ (solid lines)
and $L=2dx$ (dotted lines).
} } }

\vfill\eject

\bigskip
\centerline{\bf SUPPORTING TABLES }
\bigskip

\noindent{\bf Table~S1.}
Parameters displayed in Movie~2.

\bigskip

\settabs
\+& 123456798 & 123456789 &  \cr
\medskip
\hrule\medskip
\+& Parameters & & Definition  \cr
\medskip
\hrule\medskip
\+& $step$,   & $fburn$  & timestep, mass fraction of burnt material \cr
\+& $time$,   & $rhomax$ & time (s), maximum mass density in computational domain (g/cm$^3$) \cr
\+& \cr
\+& $fsurf$,  & $fsurf1$ & flame surface area ($^1$),  ~~~$fsurf1=fsurf/(4\pi~rfmax^2$)      \cr
\+& $rfmin$,  & $rfmax$  & min, max distance from star center to flame surface ($^2$)        \cr
\+& $ssurf$,  & $ssurf1$ & star surface area ($^1$), ~~~$ssurf1=ssurf/(4\pi~rsmax^2$)        \cr
\+& $rsmin$,  & $rsmax$  & min, max distance from star center to star surface ($^2$)         \cr
\+& $vrsmin$, & $vrsmax$ & min, max radial flow velocity at star surface (cm/s)              \cr
\+& $vrmin$,  & $vrmax$  & min, max radial flow velocity in computational domain (cm/s)      \cr
\+& \cr
\+& $etherm$  &          & thermal energy                       ($^3$) \cr
\+& $ekin$    &          & kinetic energy                       ($^3$) \cr
\+& $egrav$   &          & gravitational enegry                 ($^3$) \cr
\+& $etot$    &          & total energy = etherm + ekin - egrav ($^3$) \cr
\+& $enuc$    &          & nuclear energy                       ($^3$) \cr
\medskip \hrule
\medskip

\noindent
($^1$) scaled by $x_{max}^2$

\noindent
($^2$) scaled by $x_{max} = 5.35\times10^8$~cm.

\noindent
($^3$) scaled by $10^{50}$ergs
\vfill\eject

\bigskip
\centerline{\bf SUPPORTING REFERENCES AND NOTES }
\bigskip

\hang\noindent
S1.~W.~A.~Fowler, G.~R.~Caughlan, B.~A.~Zimmerman, {\sl Annu. Rev. Astron.
Astrophys.} {\bf 13}, 69 (1975) 

\hang\noindent
S2.~S.~E.~Woosley, W.~A.~Fowler, J.~A.~Holmes, B.~A.~Zimmerman,
{\sl Atomic Data and Nuclear Data Tables}, {\bf 22}, 371 (1978) 

\hang\noindent
S3.~F.-K.~Thielemann, M.~Arnould, J.~W.~Truran, in {\sl Advances in Nuclear Astrophysics},
E.~Vangioni-Flam, Ed. (Editions fronti\`eres, Gif-sur-Yvette, 1987), p.525

\hang\noindent
S4.~J.~W.~Truran, A.~G.~W.~Cameron, A.~Gilbert, {\sl Canadian J. of Phys.}
{\bf 44}, 563 (1966) 

\hang\noindent
S5.~D.~Bodansky, D.~D.~Clayton, W.~A.~Fowler, {\sl Astrophys. J. Suppl. Ser.}
{\bf 16}, 299 (1968) 

\hang\noindent
S6.~S.~E.~Woosley, ~W.~D.~Arnett, D.~D.~Clayton, {\sl Astrophys. J. Suppl. Ser.}
{\bf 26}, 231 (1973)

\hang\noindent
S7.~A.~M.~Khokhlov, {\sl Mon. Not. R. Astron. Soc. } {\bf 239}, 785 (1989)

\hang\noindent
S8.~V.~N.~Gamezo, J.~C.~Wheeler, ~A.~M.~Khokhlov, E.~S.~Oran,
{\sl Astrophys. J.} {\bf 512}, 827 (1999)

\hang\noindent
S9.~A.~M.~Khokhlov, {\sl Astron. Astrophys.} {\bf 245}, 114 (1991)

\hang\noindent
S10.~A.~M.~Khokhlov, http://www.arxiv.org/abs/astro-ph/0008463 (2000)

\hang\noindent
S11.~A.~M.~Khokhlov, {\sl Astrophys. J.} {\bf 449}, 695 (1995)

\hang\noindent
S12.~A.~M.~Khokhlov, {\sl J. Comput. Phys.} {\bf 143}, 519 (1998)

\hang\noindent
S13.~A.~M.~Khokhlov, E.~S.~Oran, {\sl Combust. Flame} {\bf 119}, 400 (1999)

\hang\noindent
S14.~V.~N.~Gamezo, A.~M.~Khokhlov, E.~S.~Oran, {\sl Combust. Flame}
{\bf 126}, 1810 (2001)

\hang\noindent
S15.~A.~M.~Khokhlov {\sl et al}., {\sl Astrophys. J.} {\bf 524}, L107 (1999)

\hang\noindent
S16.~D.~A.~Frank-Kamenetskii, {\sl Diffusion and Heat Transfer in Chemical
Kinetics} (Plenum, New York, 1969), chap.~6.

\hang\noindent
S17.~Ya.~B.~Zeldovich, G.~I.~Barenblatt, V.~B.~Librovich, G.~M.~Makhviladze,
{\sl The Mathematical Theory of Combustion and Explosions}
(Consultants Bureau, New York and London, 1985), chap.~4.

\hang\noindent
S18.~F.~A.~Williams, {\sl Combustion Theory} (Benjamin-Cummings, Menlo Park, 
ed.~2, 1985), chap.~5.

\hang\noindent
S19.~F.~X.~Timmes, S.~E.~Woosley, {\sl Astrophys. J.} {\bf 396}, 649 (1992)

\hang\noindent
S20.~A.~M.~Khokhlov, E.~S.~Oran, J.~C.~Wheeler, {\sl Astrophys. J.}
{\bf 478}, 678 (1997)

\hang\noindent
S21.~K.~Nomoto, D.~Sugimoto, S.~Neo, {\sl Astrophys. Space Sci.} {\bf 39}, L37 (1976)

\hang\noindent
S22.~J.~C.~Niemeyer, S.~E.~Woosley, {\sl Astrophys. J.} {\bf 475}, 740 (1997)

\hang\noindent
S23.~W.~Hillebrandt, J.~C.~Niemeyer, {\sl Annu. Rev. Astron. Astrophys.}
{\bf 38}, 191 (2000)

\hang\noindent
S24.~A.~M.~Khokhlov, E.~S.~Oran, J.~C.~Wheeler, {\sl Combust. Flame}
{\bf 105}, 28 (1996)

\hang\noindent
S25.~E.~Livne, {\sl Astrophys. J.} {\bf 406}, L17 (1993)

\hang\noindent
S26.~D.~Arnett, E.~Livne, {\sl Astrophys. J.} {\bf 427}, 315 (1994) 

\vfill\eject\end